# Microwaves reveal the nanoscale ion intercalation and edge activity of 2D catalyst

Mohamed Awadein,[†] Abhishek Kumar,[‡] Yuqing Wang,[¶] Mingdong Dong,[¶] Stefan Müllegger,[§] and Georg Gramse*,[∥]

[†]*Keysight Labs Austria, Keysight Technologies, Linz 4020, Austria*
[‡]*Institute of Semiconductor and Solid-State Physics, Johannes Kepler University, Linz 4040, Austria*
[¶]*Interdisciplinary Nanoscience Center (iNANO), Aarhus University, Aarhus C, DK-8000 Denmark*
[§]*Institute of Semiconductor and Solid State Physics, Johannes Kepler University, Linz 4040, Austria*
[∥]*Institute of Biophysics, Johannes Kepler University, Linz 4020, Austria*

E-mail: georg.gramse@jku.at

**Abstract**

The accelerated demand for electrochemical energy storage urges the need for new, sustainable, stable and lightweight materials able to store high energy densities rapidly and efficiently. Development of these functional materials requires specialized techniques that can provide a close insight into the electrochemical properties at the nanoscale. For this reason, we have introduced the electrochemical scanning microwave microscopy (EC-SMM) enabling local measurement of electrochemical properties with nanometer spatial resolution and sensitivity down to atto-Ampere electrochemical currents. The exceptional power of EC-SMM operated at radio frequency is exemplified here by the successful detection of the electrochemical activity and dynamics of molecularly thin NiCo(OH)$_2$ flakes with a spatial resolution of 16 ± 1 nm, uncovering the location of the active sites and providing atomistic details on the catalytic process that controls the electrocatalytic performance. Our results pinpoint the factors required to tune the thermodynamics of ion intercalation and to optimize the surface adsorption.

# Introduction

The development of more effective electrochemical energy storage technologies is increasingly based on functional materials structured at length scales from micro- down to nanometres.[1–3] Molecule-based materials that exhibit a precise micro- or nano-metric structural arrangement of functional molecular units have recently been reported as promising candidates.[4–6] However, researchers are still investigating most nanomaterials with bulk techniques that average over large samples rather than looking at single nanostructures with true nanoscale sensors. This is especially important for electrochemical (EC) nanoscale imaging where classical optical techniques are difficult to implement and scanning probe techniques like scanning tunneling microscopy (STM), scanning electrochemical microscopy, and other powerful variants like scanning electrochemical cell microscopy have emerged.[7] Amongst these techniques electrochemical STM,[8–10] has shown the best topographical resolution going down sub-nm and current sensitivity of sub-pA. However, the sub-pA current sensitivity is not always enough to measure electrochemical current at the nanoscale. To overcome this limitation, we have previously reported the viability of using GHz heterodyne impedance sensing method combined with EC-STM to achieve atto-Ampere (aA, $10^{-18}$ A) sensitivity.[11,12] Here, we fully exploit the potential of this so called electrochemical scanning microwave microscopy (EC-SMM) technique to resolve the ambiguity in the nature of the active site and the mechanical processes that dominate the electrocatalytic performances, exemplified with $NiCo(OH)_2$, as it is considered a storage material candidate, displaying a catalytic activity on par or better performance than other precious metal counterparts.[13,14] Previous studies on the pre-catalytic regions at various pH values demonstrated charge storage with fast kinetics, being a typical property of pseudo-capacitive materials with electrochemical (de)intercalation conversion reactions that follow thermodynamic predictions.[15,16] Herein, we report experimental measurements of EC-SMM, operated at microwave frequency (2.7 GHz) and at nanometre length scale, which reveal the local electrochemical activity induced by the edge states in nanometer thin $NiCo(OH)_2$ flakes. The results demonstrate the heterogeneity of the local intercalation kinetics and redox-processes. This information is essential for tuning the reactivity of energy materials and improve their efficiency.

## Results and Discussion

Fig. 1 a) shows a schematic diagram of the experimental setup that is able to measure local electrochemical currents down to the aA scale. At its core, it comprises a VNA integrated with an electrochemical scanning tunneling microscope EC-STM, that allows the detection of ultra-small quantities of charge-variations in the liquid and solid phase of the material below the probe. Additionally, we have a bias-T to split the DC and the microwave signal, a bi-potentiostat to maintain a constant potential between the probe and the sample while altering the electrochemical potential between the sample and the reference electrode. For imaging, a fixed bias superimposed by a small AC voltage was applied between the two working electrodes and topography-feedback was switched on while the electrochemical potential $V_{ec}$ of the working electrodes against the reference electrode can be still altered. Figure 1 b) shows the STM topography and 1 c) the electrochemical activity of a single-layered NiCo(OH)$_2$ layered double hydroxides (LDHs) sample that was prepared and adsorbed on flat substrates of highly oriented pyrolytic graphite (HOPG) as detailed in the materials and methods section. The topography image acquired in 0.1 mM KOH solution shows very flat and only 1.1 ± 0.1 nm thick flakes of approximately 400 nm in diameter, largely in agreement with previously reported AFM data.[17] Together with the topography also the voltage-dependent local electrochemical current $I_{ec}(V)$ as obtained from the measurement signals of the Vector Network Analyzer (VNA) were acquired simultaneously. An overlay of both topography and electrochemical activity is shown in Fig. 1 d). The drastic change of $I_{ec}$ by $\Delta I_{ec} = 52 \pm 4$ aA between the HOPG substrate and the NiCo(OH)$_2$ flakes clearly indicates a local increase of the electro-chemical activity. Most importantly, a pronounced increase of $I_{ec}$ at the borders of the flake can be observed, which is clearly uncorrelated with topography and not related to a feedback artifact (details in Supporting Note 1). Note the high sensitivity of the $I_{ec}(V)$ image corresponding to aA and its lateral spatial resolution $\Delta x_{I_{ec}} = 16\pm1$ nm, which is almost as good as the STM imaging resolution $\Delta x_{STM} = 10\pm1$ nm in this image.

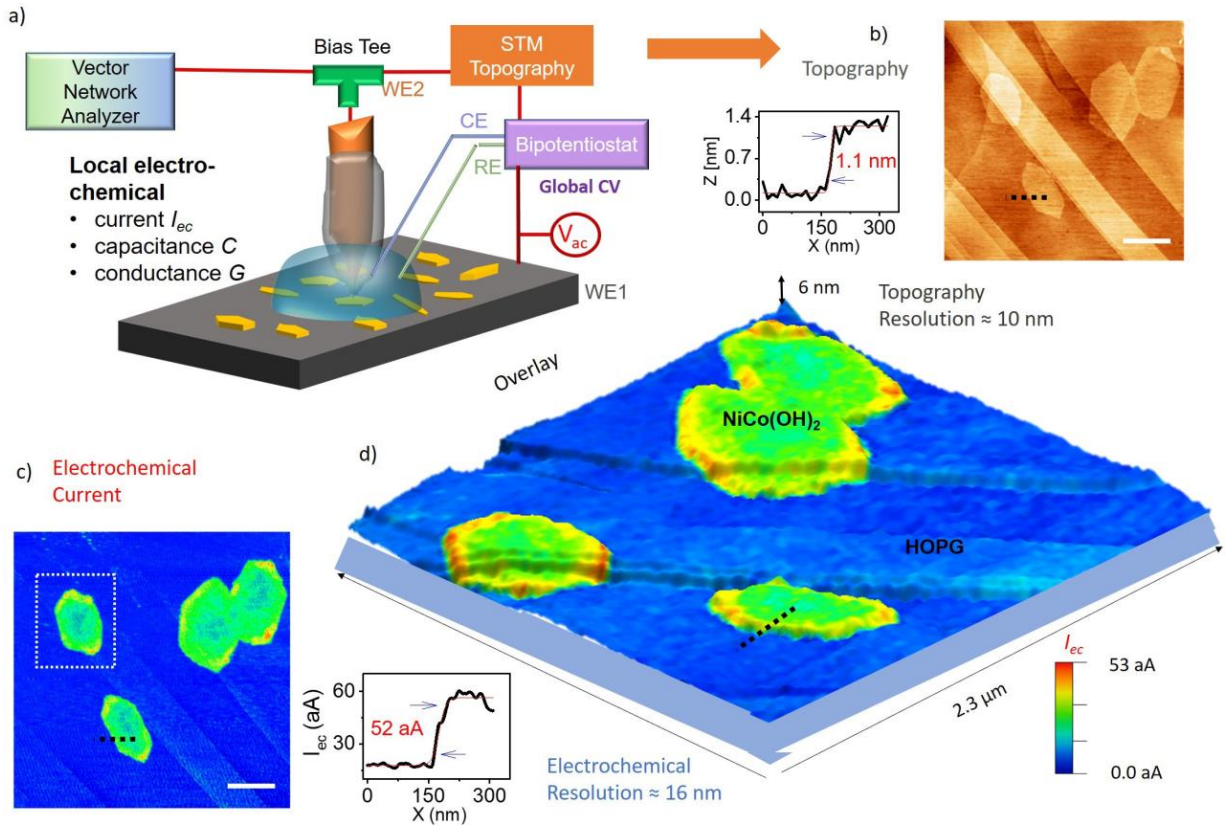

*Figure 1: Local electrochemical activity of NiCo layered double hydroxides flakes resolved with nm resolution by the electrochemical SMM. a) Schematic diagram of the electrochemical SMM setup scanning HOPG sample (grey colour) with NiCo layered double hydroxides flakes (yellow colour). b)-d) STM topography, electrochemical current acquired by the SMM and overlay of electrochemical current and topography, respectively. Profile lines of topography and ec-current across the black dashed line is shown ($V_{ec}$=480 mV, modulation frequency is 1 kHz). Scale bar is 300 nm.*

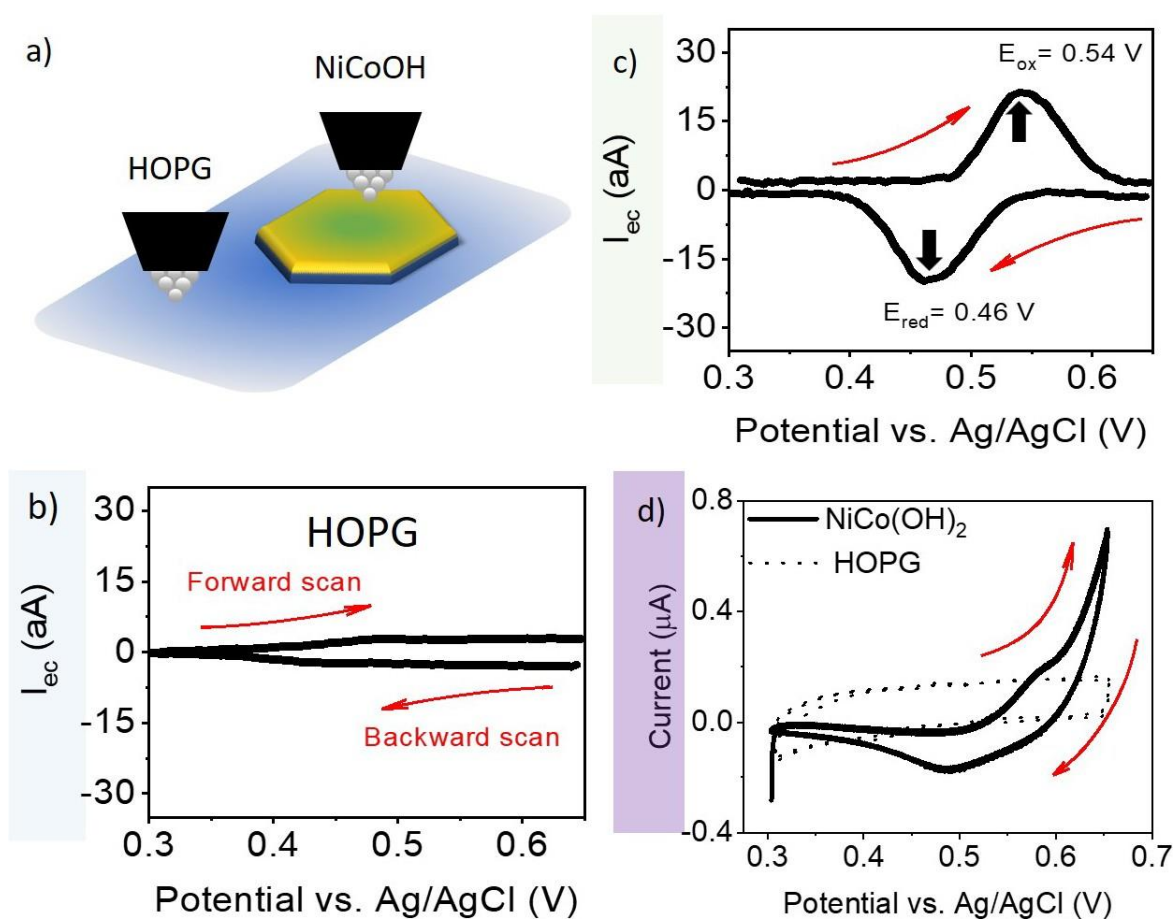

*Figure 2: Localized cyclic voltammogram (CV). a) Sketch shows the localized measurements. b) Obtained on the bare HOPG substrate, c) on NiCo layered double hydroxides flakes, respectively. d) Global CV on HOPG substrate with NiCo layered double hydroxides flakes (solid line) and pristine HOPG (dashed line). The CVs were measured, in a three-electrode configuration, with respect to a Ag/AgCl reference electrode, at a scan rate of 10 mV/s.*

The electrochemical activity visualized in Fig. 1 depends strongly on the electrochemical potential, $V_{ec}$, applied to the sample. Besides imaging at fixed $V_{ec}$, we show in Fig. 2 that the

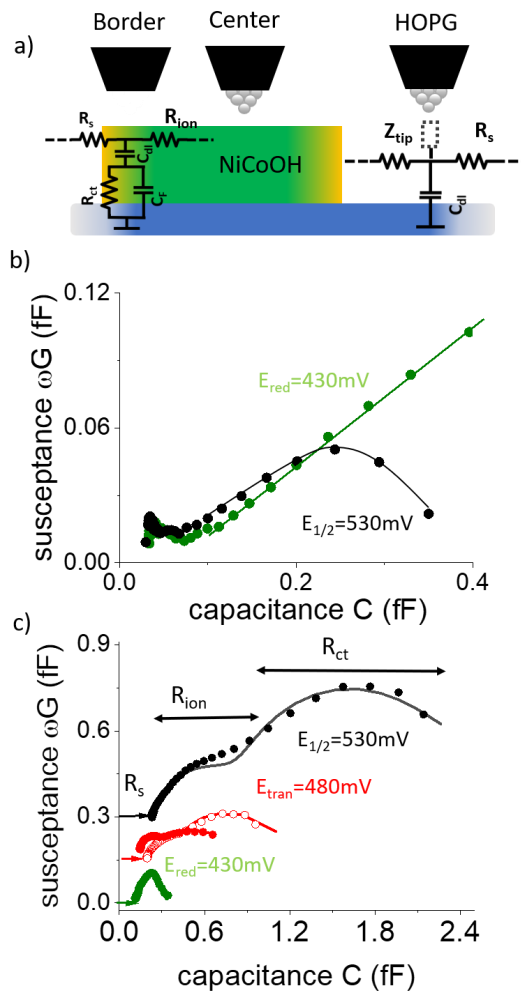

*Figure 3: Local electrochemical reaction kinetics. a) The tip-sample interaction modeled by equivalent circuits considering faradic reactions ($R_{ct}$, $C_f$), diffusive contributions ($R_{on}$, $R_s$, $C_{dl}$). b) LEIS of HOPG and at 430 mV and 530 mV, the frequency spectrum ranges from 100 Hz to 100 kHz. c) LEIS of NiCo layered double hydroxides at the edge (curve B) and near the centre (C) of the NiCo layered double hydroxides flake at 480 mV and 530 mV the spectra have two relaxation times around 1.3 kHz (faradic) and 27 kHz (non-faradic), and single relaxation of 25 kHz at 430 mV. Solid lines show the fitting of the equivalent circuits to the data.*

microscope allows us to acquire local cyclic voltammograms (LCVs) directly over the bare HOPG and the flakes, respectively. As expected, the LCV on the HOPG (Fig. 2b) shows no redox reaction and besides the capacitive charging of the double layer,[18] the signal stays almost constant with varying $V_{ec}$. For comparison we acquired in Fig. 2 d) conventional CV on a 10 mm$^2$ large sample of the pristine HOPG (dashed curve). Here as well only a capacitive behavior due to double layer charging is observed. The conventional CV of

NiCo(OH)$_2$ on HOPG (solid curve) in Fig. 2 d) shows a reversible oxidation activity of:

$$Co^{2+}(OH)_2 \rightarrow Co_3^{2.66+}O_4 \qquad (1)$$

with a strongly broadened oxidation peak at 580 ± 2 mV and a reduction peak at 490 ± 2 mV with half-wave potential $E_{1/2}$ = 535 ± 2 mV, in agreement with literature.[17,19] Notice that the oxidation and reduction peaks are much more clearly visible in the measured LCV curve (c) extracted on the NiCo(OH)$_2$ flake than in the global CV curve (d). Since the scanning probe used here exhibits a tip apex radius of ≈ 10 nm, it detects the local electrochemical activity of a very small surface area of the NiCo(OH)$_2$.

The positions of the oxidation and reduction potentials obtained from the localized CV are offset by −30 ± 5 mV with respect to the global ones. We attribute this offset to the difference in potential $V_{bias}$ = −70 mV between the probe and the substrate that is required for the tunneling feedback. Most notably is the observed heterogeneity of $I_{ec}$ across the individual flakes showing increased electrochemical activity at the edges than in the centre of the flakes. This can be either associated with low-dimensional electronic edge states and thus with a locally varying electrochemical potential as found in transition metal dichalcogenides like MoS$_2$ and other 2D-materials[20,21] and/or with diffusion limited processes as predicted recently by T.J. Mefford, et al.[16] To distinguish between diffusion processes and electrochemical charge transfer processes taking place below the probe, we have measured the dynamics of the electrochemical properties by operating the EC-SMM in the localized electrochemical impedance spectroscopy (LEIS) mode.

**Local dynamics**

The EC-SMM acquires the localized capacitive and conduction-related electric currents induced by a small time-dependent (harmonic) electric potential $V_{ac}(t)$ oscillating at frequency $f$ around the static electrochemical potential $V_{ec}$ of the sample in a frequency-dependent mea-surement mode. Similar to conventional EIS measurements,[22,23] acquisition of the frequency dependent complex impedance allows to assess the ion dynamics in the solution and the kinetics of the electrochemical charge transfer process. This is achieved by numerical fitting of the experimental data to an equivalent circuit model (see Fig. 3 a) – herein achieved with nanometre spatial resolution on specific areas of the sample selected by the position of the scanning probe. Figures 3 b) and c) show the LEIS acquired on NiCo(OH)$_2$ (b) and on the bare HOPG substrate (c) at different potential values of $V_{ec}$ = 430 mV, 480 mV and 530 mV, respectively; these values lie close to the values of $E_{red}$, $E_{tran}$ and $E_{1/2}$ determined above. The LEIS data acquired on the bare substrate (Fig.3 b) shows two relevant features. One at the high-frequency end of the spectrum, close to the origin of the Nyquist plot that is associated with the probe impedance (dashed $Z_{tip}$ element). At the low frequency end the straight line (430 mV) and the semi-circle (530 mV) are characteristic for a semi-infinte or finite diffusion processes, respectively, that can be modelled by a Warburg element or the distributed model depicted which includes solution resistance $R_{s,t}$ and diffuse layer capacitance $C_{dl}$.[18] Through numerical fitting of the equivalent circuit models to the experimental data we extracted a respective diffusion coefficient $D_{OH^-,K^+}$ = (5.8 ± 1) · 10$^{-9}$ m²/s in the electrolyte.

For the LEIS measurement on NiCo(OH)$_2$, shown in Fig. 3 c), we find again a clear presence of the ionic diffusion process that is characterized by solution resistances $R_s$, $R_{ion}$ outside and inside the flake, respectively, and the local double layer capacitance $C_{dl}$. In the frequency range of $f$ = 0.1−5 kHz, where we had seen a diffusion related Warburg impedance before, we observe additional contributions, which are related to the actual electrochemical charge transfer resistance $R_{ct}$ as well as to the increasing incorporation of the OH$^-$ ions into the flakes causing the buildup of a Faradaic capacitance $C_F$.

While for a potential value of $V_{ec}$ = 430 mV, which lies well below $E_{1/2}$, the LEIS on NiCo(OH)$_2$ shows no Faradic reaction, for $V_{ec}$ = 480 mV both a Faradic and non-Faradic components are clearly visible at 1.3 kHz and 30 kHz, respectively. The LEIS at the edge of the NiCo(OH)$_2$ flake (red hollow circles) shows about two times larger $C_F$ and approximately 20% lower $R_{ct}$ than the LEIS acquired at the center of the flake (fitting results are summarized in Table 1). Since $R_{ct}$ depends on the electrochemical potential, our result points to a local increase of $V_{ec}$ on the flake boundaries with respect to the flake center. Increasing further the potential to 530 mV, the LEIS spectrum of NiCo(OH)$_2$ shows again an increase of faradic and non-faradic contributions ($C_F$ doubles and $R_{ct}$ halfs) with respect to the values obtained on the edges at 480 mV. We remark that LEIS allows us to quantitatively distinguish between the $R_{ct}$ and the mass transport – which cannot be separated by a conventional cyclic voltametry measurement.

Table 1: Equivalent circuit parameters extracted from fitting shown in Fig. 3.

| $V_{ec}$ edge | 430 mV | 480 mV | 530 mV |
|---|---|---|---|
| Position | center | center / edge | center |
| $R_{sol}(\Omega/cm^2)$ | 0.11 ± 0.01 | 0.14 ± 0.01 / 0.18 ± 0.01 | 0.20 ± 0.02 |
| $R_{ct}(\Omega/cm^2)$ | 28 ± 3 | 16 ± 2 / 13 ± 3 | 6 ± 1 |
| $R_{ion}(\Omega/cm^2)$ | 0.18 ± 0.02 | 0.44 ± 0.02 / 0.71 ± 0.03 | 1.78 ± 0.05 |
| $C_f(\mu F/cm^2)$ | 16 ± 2 | 54 ± 3 / 160 ± 6 | 510 ± 16 |
| $C_{dl}(\mu F/cm^2)$ | 17 ± 3 | 165 ± 15 / 50 ± 4 | 16 ± 2 |
| | HOPG | | |
| $D_{OH^-, K^+}(m^2/s)$ | 5.8 · 10$^{-9}$ | – | 4.8 · 10$^{-9}$ |

## Local evolution of chemical intercalation reaction

We found from the LEIS measurements that $C_F$ and $R_{ct}$ are clearly changing with the applied electrochemical potential and a significant change in $C_F$ was observed at different locations of the flake. The local values of the respective resistances and capacitances are compiled in Table 1. In order to systematically visualize the position-dependent (local) electrochemical properties across individual flakes, we image the individual flakes at increasing electrochem-ical potentials and a fixed modulation voltage of 1 kHz and applied to the substrate. The STM topography image in Fig. 4 a) was acquired at $V_{ec} = 430$ mV simultaneously with an electrochemical image (i.e. recording the position-dependent $I_{ec}$) and it reveals that the flake is rather flat and ca. 1 nm high; its shape is hexagonal-like with a diameter of ca. 400 nm.

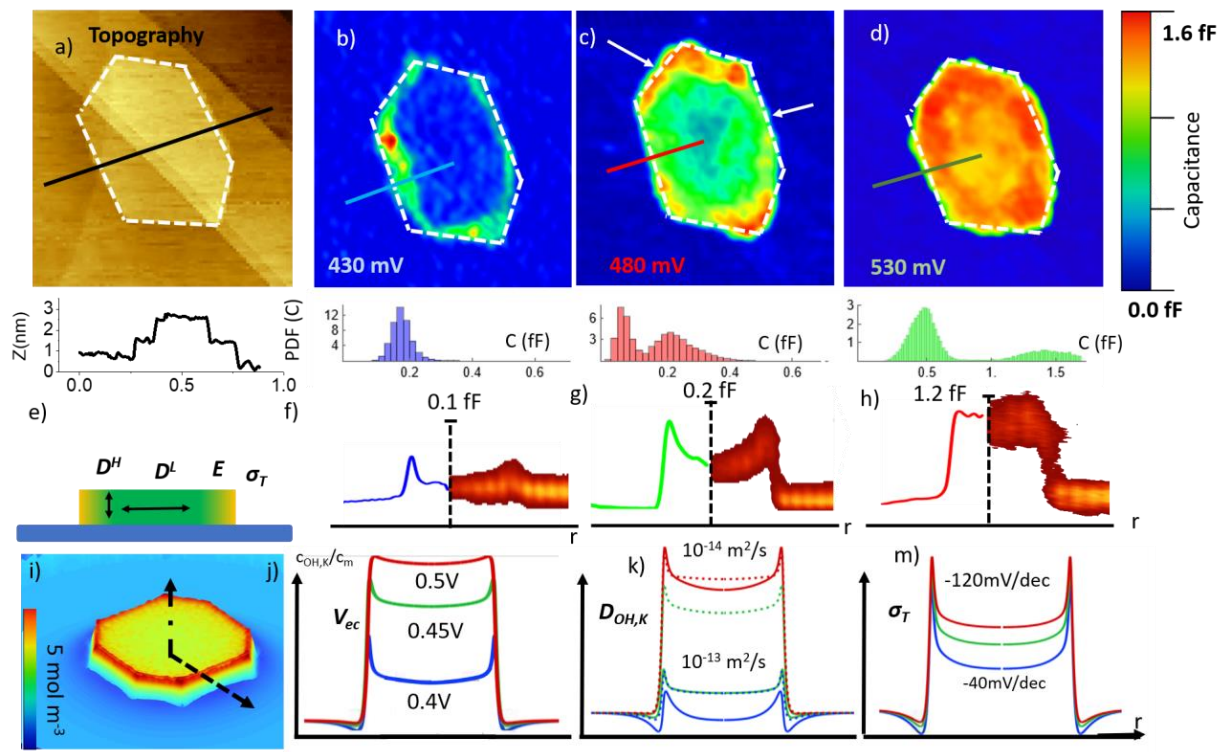

*Figure 4: Local activity, charge intercalation and diffusion processes. a) Topography of single NiCo layered double hydroxide flake on HOPG substrate. b)- d) Electrochemical capacitance images $\Delta C$ acquired at $V_{ec}$ 430 mV, 480 mV and 530 mV including histograms ( modulation frequencies are 1 kHz and 2.7GHz). e) Molecular model of NiCo layered double hydroxides flakes on a HOPG with their active sites. f-h) Statistical distribution of obtained capacitance data locally resolved at increasing distance from the flake center as indicated by the solid line in b)-d). Capacitance mean value (left) and distribution data (right) are shown. i) Ion concentration distribution (normalized $c/c_{max}$ are shown) simulated in 3D by finite element modeling ($\omega_{ac} = 1 kHz$, $V_{ec} = 0.5$, $D_s = 10^{-13} m^2/s$, if not otherwise specified, details in the Methods section). j) Modelled concentration cross-section for increasing electrochemical potential $V_{ec}$ applied to the substrate. k) Modelled capacitance cross-section for different values of diffusion constant in flake ($D_{s,L/V}$). Solid lines show identical lateral $D_{s,L}$ and vertical $D_{s,V}$ diffusion, dashed lines $D_{s,L} = 10 \cdot D_{s,V}$. m) Modelled effect of Tafel slope σ.*

Figs. 4 b)- d) show electrochemical capacitance images of the individual flake recorded at three different potential values ranging from the fully reduced state at $V_{ec,ox} = 430$ mV, to a slightly higher potential of 480 mV and finally to a potential close to the half-wave potential $E_{1/2} = 530$ mV; the local change of the capacitance $\Delta C$ with respect to the pristine HOPG substrate is displayed in units of fF and plotted in a respective color code for clarity. Notice that the local value of $\Delta C$ is directly related to the amount of intercalated electric charge in the sample at the position of the STM tip, resulting from the (local) electrochemical reaction. The insets underneath each capacity image show histograms of the $\Delta C$ values, revealing an increase of the average capacitance on the flake with increasing electrochemical potential. However, since the histograms show a superposition of all locations on the sample, we analyzed the statistical distribution of the capacitance value in dependency of the distance from the flake center. These local distribution which are shown in Figs. 4 f)-h) and follow a log-normal distribution from which we have determined a mean value $\mu$ and a standard deviation $\sigma$. At the two lower potentials the calculated $\mu$ values displayed in Fig. 4 h) show the increase of the local capacitance at the edges, however, this is almost not present anymore at 530 $mV$. At 480 $mV$ we also note, that some edges of the flake show a higher activity than the others. Despite the generally higher capacitance at the borders we find that the $OH^-$ ions finally also intercalate in the pockets and wrinkles, that are local imperfections, throughout the flake.

## Discussion

The observed increase of the electro-chemical activity at the borders of the flakes can be due to a number of reasons. First of all, it depends on the exchange current density or $R_{ct}$ that is a function of the applied potential, which should be high enough to overcome the energy barrier of the reaction i.e. the oxidation potential. A higher oxidation potential of the flake interior than on the edges could be therefore related with the observed results. It was argued that this difference in the work function may also lead to a space charge layer between the interfaces of the facets[15,24] resulting in a decrease of the local binding energy and making those molecules near the edges more weakly bound than the interior molecules. Secondly, the intercalation rate depends also on the availability of the $OH^-$ ions and the access to the reaction sites, which is mainly facilitated at the edge facets. [25,26] In context with this restricted ion movement in the <001> direction in the absence of extended defects in the $CoO_2$ layers, could prevent the basal planes from serving as reaction sites to initiate the bulk redox transformation reactions. Consequently, intercalation could be heavily dependent on the in-plane diffusion processes of the ions in the flake.[25,26] To clarify the actual origin of the local charge dynamics on the flakes observed in our measurements we carried out finite element simulations and compared them with the measurements. The model in Fig. 4e) considered for this resembles the experimental conditions and we study the effect of the diffusion constant $D_s = D_{OH^-,K^+}$ in the flake, the applied potential $V_{ec}$, the local oxidation potential $E_{1/2}$ and the Tafel slope $\sigma_T$ on the simulation results. By reducing the diffusion constant in the vertical direction we can simulate reduced access of the ions to the reaction sites. We simulated the ion distributions and electro-chemical potential for the measured sample and calculated a local ion concentration $c_{OH,K}$ that is proportional to the electro-chemical capacitance $C \propto c_{OH,K}$. We display the result in Fig. 4i for the three-dimensional simulation of the hexagonal flake. Details on the calculation can be found in the materials and methods section. As can be seen the edges of the flake show a clearly increased $c_{OH,K}$ signal with respect to the centre even in absence of any locality effect like restricted diffusion etc. We note that the decreased diffusion constant in the flake leads to an approximately quadratic capacitance profile, however, the interplay between migration and diffusion is responsible for the specific charge distribution in the flake. A more qualitative comparison with the experimental data is possible from the simulated cross-sections of $C$ showing in Fig. 4j)-m) the effect of the applied $V_{ec}$, diffusion constant $D_s$ and an increase of $\sigma_T$ on the $c$ profile lines, respectively. As can be seen in Fig. 4 the applied potential has a direct impact on the overall signal difference between substrate and flake and also the edge effect is observed for low $V_{ec}$. Similarly, an increase of the Tafel slope leads to a more pronounced edge effect. The decrease of the diffusion constant in the flake by $\approx$ 4-5 orders of magnitude $D_s \approx 10^{-14} m^2/s$ with respect to the bulk solution would be in line with the observed capacitance profile. We note that for $D_s > 10^{-14} m^2/s$ mainly the lateral diffusion is relevant, only for diffusion constants below this value we find that restricted vertical diffusion leads to less variation throughout the flake, but steep increase at the edges. This supports the picture that the missing access of ions in the center could be the reason for the more pronounced intercalation process at the borders of the flake. As such, imperfections in the lattice structure can also enhance or decrease the activity as we can observe when comparing the right edge showing a lower activity with the top and bottom edges. A local increase of $E_{1/2}$ other than through the Nernst equation would be possible, but is not required to explain the observed results.

## Conclusion

In this paper, we have exploited the unprecedented resolution and sensitivity of EC-SMM to reveal the local chemical dynamics of ion intercalation in nanostructured materials for energy storage by local electrochemical impedance spectroscopy. The investigated flakes of NiCo layered double hydroxides store energy in the intercalation process and are potential candidates for catalytic OHR. We were able to locally quantify for the first time the reaction dynamics of the intercalation process at the nanoscale, where we found the active site of the molecule concentrates at the edge facets, which is also responsible for the redox reaction. The measured capacitance at the edge was higher than at the centre of the flakes. We resolved the process of the charge intercalation along the 500 nm wide flake and found that the intercalation between the layer starts from the edge towards the middle in a slow diffusion process. These results suggest that the capacity of energy storage material can be improved by tuning the surface and the thermodynamics of the molecule. Also, the systematic inclusion of lattice imperfections could improve the effective intercalation rate.

## Materials and Methods

**Electrochemical materials** Nickel nitrate $Ni(NO_3)_2 \cdot 6H_2O$, 98% cobalt nitrate ($Co(NO_3)_2 \cdot 6H_2O$, 98%), hexamethylenetetramine (HMT) and formamide were purchased from Alfa Aesar.

Sodium dodecyl sulfate (SDS) was purchased from Fisher Scientific.

**Synthesis of single layered-NiCo LDHs.** Bulk NiCo LDHs were prepared via a hydrothermal procedure. 0.5 M Ni(NO$_3$)$_2$.6H$_2$O (3 mL), 0.5 M Co(NO$_3$)$_2$.6H$_2$O (1 mL), 0.25 M SDS (40 mL) and 1 M HMT (12 mL) solutions were mixed in a 200 mL Teflon vessel with 44 mL deionized Milli-Q water. The reaction was kept at 120 C° for 24 h, then cooled down to room temperature. The precipitates were washed with water and ethanol several times and dried overnight. Next, 15 mg bulk NiCo LDHs were dispersed in 30 mL formamide and maintained at 40°C for 120 h. The suspension was centrifuged at 2,000 rpm for 0.5 h, and the as-obtained supernatant was centrifuged at 10,000 rpm for 0.5 h. Then the sediments were washed with water and ethanol several times (13,000 rpm for 0.5 h, and the product was suspended in 2 mL of ethanol. The solution was stored at 4 °C in a refrigerator as a dispersion medium. The same dispersion medium was also used during transportation.

**Sample preparations** HOPG ZYB from TipsNano (15 mm × 15 mm × 2 mm) was used as a substrate. First, a scotch tape exfoliation was performed to peel away the top layers of HOPG. Then, 10 uL of *NiCo* layered double hydroxides dispersion was drop-casted onto the freshly cleaved HOPG surface. It was allowed to dry and annealed at 180 °C for 60 min on a hot plate inside a fume hood. Typical dimensions of the flakes prepared by this method give a cross-sectional diameter of (0.5 ± 0.2) *µ*m and thickness 1 ± 0.1nm.

**Microwave-Electrochemical Microscopy** A Keysight 5400 AFM mount connected to a STM scanner was used to perform tunneling microscopy. The Keysight controller supplied the DC signal (for eg. 50 mV, 70 pA) required for tunneling. The tunneling cable on the tip side was additionally connected to a Vector Network Analyzer (VNA) through a bias tee. This connection scheme allowed VNA as a radio frequency source to send RF signal (for eg. 2.7 GHz, -10 dBm) to the tip, and to collect the reflected $S_{11}$ signal back to VNA. An impedance tuner (Maury Microwave 8420) was used for tuning the impedance of the RF branch. This enabled us to minimize the reflected signal $S_{11}$ at the frequency of choice. A lock-in amplifier was used to achieve a better signal-to-noise ratio, giving the amplitude and phase of d$S_{11}$/dV as output. For electrochemical measurements, a potentiostat was connected and used in a four-probe configuration: NiCo layered double hydroxides on HOPG as a working electrode, a platinum wire (0.3 mm diameter, 5 cm long) as a counter electrode, an in-house coated Ag/AgCl wire (0.5 mm diameter, 5 cm long) as a reference electrode and the wax coated Pt-Ir tip as the fourth probe. A 1 mM solution of *KOH* in water was used as an electrolyte, which was bubbled with nitrogen gas for 10 minutes immediately before performing cyclic-voltammetry measurements. For the coated Pt-Ir STM tips Apiezon wax was melted and a Pt-Ir STM tip was quickly passed across it to prepare an STM tip with only a few nm opening at the tip apex but otherwise electrically insulated (details in[11]). All measurements presented here were performed at room temperature.

**Finite element modelling** Comsol Multi-Physics 6.1, electrochemistry module, was used to simulate the electrochemical response of the flake. Full 3D simulations and 2D axially symmetric simulations were carried out calculating the tertiary current distribution in steady-state, harmonic perturbation, in a 1$\mu m^2$x1$\mu m^2$ domain containing the KOH electrolyte. The flake was modelled as a porous electrode of 500 nm in diameter and 2 nm in height. The diffusion constant, Tafel slope, applied electrochemical potential and perturbation frequency were set as parameters. The equilibrium potential of the reaction was set to 500 mV and it was assumed that it changes following the Nernst equation. Very fine meshing (2 nm, in the electrode area) including boundary mesh was required to converge the model.

# Acknowledgement

We thank the European Research Council for funding the project SARF *Single Atomic Radio frequency Fingerprinting* (Grant Agreement No. 771193) through an ERC Consolidator Grant. We thank the European Innovation Council for funding the project 4D-NMR under grant agreement No 101099676 through the European Union's Horizon Europe research and innovation program. Additional support from the NanoBat project funded by the EU's

H2020 research and innovation program under Grant Agreements no. 861962, the ATTRACT 2 project UNICORN funded by European Commission under Grant Agreement 101004462 and from the MSCA-ITN project BORGES under grant agreement 813863 is gratefully acknowledged.

# Supplementary Note 1

## Imaginary part of $C*$ and $S_{11}$ images

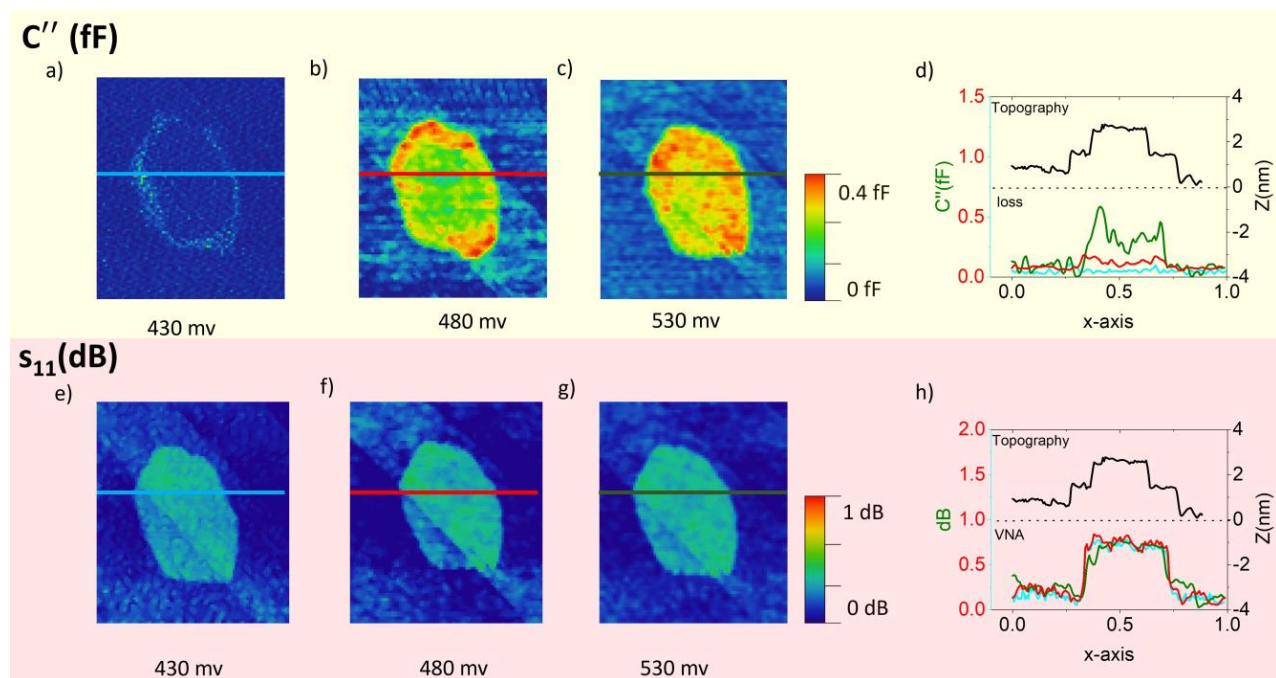

*Figure S1: Imaginary part of the acquired complex capacitance $C*$ and $S_{11}$ image a-c) $C'$ at different potentials as indicated. d) Topography profile in comparison with $C'$ profiles. e-g) Images of $S_{11}$ amplitude at different potentials. h) The profile of topography versus the profile of $S_{11}$.*

The images of the $\omega G = C''$ channel at different potentials show similar behavior as the images of $C$. Where the edge shows enhanced electrochemical activity as well. The cross-talk can be also verified here by comparing the profiles. Figure S1 c) shows the images of $S_{11}$ at a different potential. Nevertheless, there is almost no potential dependency on the $S_{11}$ profiles. In contrast to direct $S_{11}$ images (Fig. S1) that visualizes both the geometrical capacitance of the flake together with the presence of surface charges, in $\frac{dS_{11}}{dV}$ images only the charge that follows the kHz modulation voltage is measured. The cross-talk can be also verified from the HOPG step that can hardly be distinguished by the $S_{11}$ signal.

# Supplementary Note 2

## Morphology and Topography

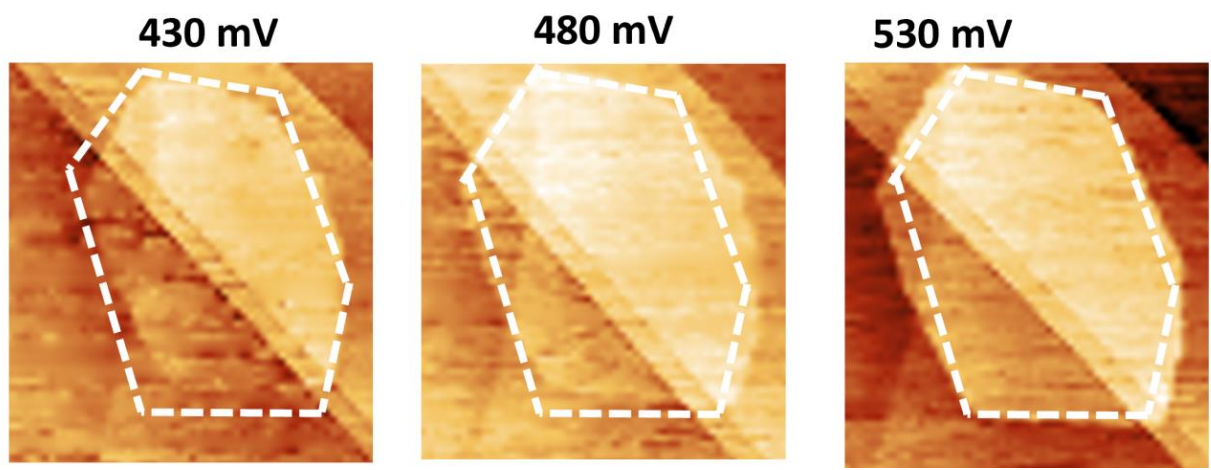

*Figure S2: Topography of a NiCo layered double hydroxides flakes in 0.1 mM KOH at different applied voltages.*

The size and the topographical morphology of the flake changed at different electrochemical potentials.